\documentclass[10pt,aps,prc,twocolumn,showpacs,preprintnumbers,amsmath,amssymb]{revtex4-1}

\usepackage{graphicx,psfrag}

\usepackage{color}

\definecolor{purple}{rgb}{0.6,0,0.5}

\definecolor{verde}{rgb}{0,0.5,0}

\usepackage[colorlinks=true, pdfstartview=FitV, linkcolor=blue,
citecolor=blue,urlcolor=navyblue]{hyperref}

\begin{document}

\title{Correlations between critical parameters and bulk properties of nuclear matter.} 

\author{O. Louren\c{c}o$^1$, B. M. Santos$^2$, M. Dutra$^3$, and A. Delfino$^2$}

\affiliation{$^1$Universidade Federal do Rio de Janeiro, 27930-560, Maca\'e, RJ, Brazil \\
$^2$Instituto de F\'isica, Universidade Federal Fluminense, 24210-150, Niter\'oi, RJ, Brazil  \\
$^3$Departamento de Ci\^encias da Natureza, IHS, Universidade Federal Fluminense, 28895-532, Rio 
das Ostras, RJ, Brazil}

\date{\today}


\begin{abstract}

The present work starts by providing a clear identification of correlations between critical 
parameters ($T_c$, $P_c$, $\rho_c$) and bulk quantities at zero temperature of 
relativistic mean-field models (RMF) presenting third and fourth order self-interactions 
in the scalar field $\sigma$. Motivated by the nonrelativistic version of this RMF model, 
we show that effective nucleon mass ($M^*$) and incompressibility ($K_o$), at the 
saturation density, are correlated  with $T_c$, $P_c$, and $\rho_c$, as well as, binding 
energy and saturation density itself. We verify agreement of results with previous 
theoretical ones regarding different hadronic models. Concerning recent experimental data 
of the symmetric nuclear matter critical parameters, our study allows a prediction of 
$T_c$, $P_c$ and $\rho_c$ compatible with such values, by combining them, through the 
correlations found, with previous constraints related to $M^*$ and $K_o$. An improved RMF 
parametrization, that better agrees with experimental values for $T_c$, is also indicated.

\end{abstract}

\maketitle

\section{Introduction}

One of the most successfully methods to treat strongly interacting matter at the hadronic level is 
QHD (Quantum HadroDynamics)~\cite{qhd}. In this quantum field theory, that adequately 
incorporates effects of quantum mechanics and relativity, nucleons are described by the Dirac spinor 
$\psi$, and the exchanged mesons by $\sigma$ and $\omega$ fields, responsible to take into 
account the attractive and repulsive nature, respectively, of nuclear interaction. The 
nuclear saturation is obtained in this model by the near cancellation of the scalar and 
vector potentials, written in terms of the $\sigma$ and~$\omega$ mean-field values. By 
using this type of treatment, many effective models have been constructed in order to 
better describe infinite nuclear matter and finite nuclei properties. The starting model 
was initially developed by Walecka in the seventies~\cite{walecka}. This seminal work was followed by 
many other improved versions, and several variations (parametrizations) were built. For a collection 
of such models, see for instance, Refs.~\cite{rmf,bali1}.

Concerning these particular relativistic mean-field (RMF) models, a specific and detailed 
study on the possible correlations presented by the bulk parameters they describe, and how (under what 
conditions) they can emerge, has not yet been completed. Many investigations were 
performed showing indications or trends of correlations, but clear conditions on what 
physical parameters are important to generate such trends are not totally established. In 
that direction, we have developed investigations on the subject in Ref.~\cite{bianca}, and 
verified that, for the RMF model described by the Lagrangian density presenting only the 
$\sigma^3$ and $\sigma^4$ terms for scalar meson self-interaction, the effective nucleon 
mass plays an important role on the arising of correlations between bulk quantities in 
both isoscalar and isovector sectors. For the latter sector, for instance, we have 
shown~\cite{bianca} how symmetry energy~\cite{bali2,bali3,bali4,bali5,bali6,fred} 
correlates with its next order bulk parameters, namely, slope and curvature. In this work, 
we proceed to further investigate these possible correlations, but now analyzing the 
finite temperature regime of the RMF model presenting $\sigma^3$ and $\sigma^4$ 
self-interactions. We study here how the critical parameters ($T_c$, $P_c$, $\rho_c$) of 
this model can correlate with zero temperature bulk quantities, such as effective mass and 
incompressibility. In order to perform such an analysis, we first use the analytical 
structure of the nonrelativistic version of this RMF model to predict correlations of 
$T_c$, $P_c$ and $\rho_c$ with bulk parameters at $T=0$. The details of these 
calculations are presented in Sec.~\ref{nrl-analysis}. In Sec.~\ref{rmf-analysis}, we show 
how correlations of RMF model emerges, motivated by results presented in the previous 
section. We also compare our findings with theoretical results of former 
investigations~\cite{kapusta,lattimer,natowitz,rios}, and with available experimental 
values concerning $T_c$, $P_c$ and $\rho_c$ of infinite symmetric nuclear 
matter~\cite{natowitz,karn1,karn2,karn3,karn4,karn5,elliott}. We show which 
parametrizations are compatible with experimental data, by combining the critical 
parameters values with other bulk parameters constraints, such as one related to the 
effective nucleon mass. Finally, in Sec.~\ref{summ-concl}, we provide a summary and the 
main conclusions of our work.

\section{Nonrelativistic analysis} 

\label{nrl-analysis}

In order to analyze possible correlations of the critical density, temperature and 
pressure of neutron-proton symmetric nuclear matter, we investigate a particular 
nonrelativistic model, namely, the nonrelativistic limit (NRL) of the relativistic 
nonlinear point-coupling (zero range) model with self-interactions in the 
$\bar{\psi}\psi$ condensate until fourth order. As pointed out in previous 
studies~\cite{bianca}, the model generated from this NRL exhibits many explicit 
correlations among zero temperature bulk quantities, and can also be used as a starting 
point to search the same correlations in finite range RMF parametrizations presenting 
self-interactions in the scalar field ($\sigma$), also until fourth order. In the 
following, we present the formalism and construction of the main equations of state of 
this NRL model.

\subsection{Formalism at zero temperature}

In nuclear physics, point-coupling (or zero range) models assume that nucleons interact 
with each other only when they are in contact - a zero interaction 
range means there is no meson exchanges between protons and neutrons. From a 
qualitative point of view, since the nuclear interaction range is inversely proportional 
to the mesons mass, one can consider a point-coupling model as one in which the mesons 
mass are high enough (infinity), leading to a vanishing nuclear range.

In nonrelativistic frameworks, the most known and used point-coupling model is the Skyrme 
one~\cite{skyrme}, successfully used in description of infinite nuclear matter and finite 
nuclei. In relativistic contexts, on the other hand, nonlinear relativistic point-coupling 
(NLPC) models have been 
applied~\cite{nlpc2,nlpc3,nlpc4,newrefpc1,newrefpc2,nlpc1,sulaksono,pclambda} to extract nuclear 
ground-state observables, with results comparable in quality to those obtained by usual 
relativistic finite range models. Here, we use the point-coupling version of the 
finite-range RMF model presenting terms of $\sigma^3$ and $\sigma^4$ (we discuss this 
particular model in the next section). Its Lagrangian density, for symmetric neutron-proton 
system and in the zero temperature regime, is given by

\begin{eqnarray}
\mathcal{L}_{\mbox{\tiny NLPC}}&=&\bar{\psi}(i\gamma^{\mu}\partial_{\mu}-M)\psi
-\frac{1}{2}G^{2}_{\mbox{\tiny V}} 
(\bar{\psi}\gamma^{\mu}\psi)^{2}+\frac{1}{2}G^{2}_{\mbox{ \tiny S}}(\bar{\psi}\psi)^{2} 
\nonumber \\
&+&\frac{A}{3}(\bar{\psi}\psi)^{3}+\frac{B}{4}(\bar{\psi}\psi)^{4}.
\label{nlpc-lag}
\end{eqnarray}

The Euler-Lagrange equation applied to $\bar{\psi}$ in Eq.~(\ref{nlpc-lag}) gives rise to 
the following Dirac equation for the $\psi$ field,

\begin{eqnarray}
(i\gamma^{\mu}\partial_{\mu} - M + G^{2}_{\mbox{\tiny S}}\rho_s -
\gamma^0G^{2}_{\mbox{\tiny V}}\rho + A\rho_s^2 + B\rho_s^3)\psi = 0, \quad\,\,\,
\label{nlpc-dirac}
\end{eqnarray}

with $\rho_s=\bar{\psi}\psi$. Here, $\rho$ is the nucleon density. The nonrelativistic 
limit of the NLPC model~\cite{sulaksono} is then obtained by first writing the large 
component~$\phi$ of the Dirac field~$\psi$ in terms of the small one~$\chi$. This 
procedure leads to

\begin{eqnarray}
(\boldsymbol{\sigma\cdot k}\,\tilde{B}\,\boldsymbol{\sigma\cdot k}\,+\,M+S+V)\phi &=& 
E \phi
\label{nlpc-large}
\end{eqnarray}

with

\begin{eqnarray}
\tilde{B} &=& \frac{\tilde{B}_0}{1 + (\epsilon-S-V)\tilde{B}_0} \, \simeq \, \tilde{B}_0 
+ \tilde{B}_0^2(S+V-\epsilon),
\label{nlpc-exp}
\end{eqnarray} 

being $\tilde{B}_0 = 1/[2(M+S)]$, and $\epsilon = E - M$. The vector and scalar
potentials are, respectively, $V=G^{2}_{\mbox{\tiny V}}\rho$ and 
\mbox{$S=-G^{2}_{\mbox{\tiny S}}\rho_s-A\rho_s^2-B\rho_s^3$}. By using in 
Eq.~(\ref{nlpc-large}) the approximation~(\ref{nlpc-exp}), and taking into account an 
expansion up to order $(k/M)^2$, one can derive the following single-particle energy,

\begin{eqnarray}
H &=& \frac{k^2}{2M^*} + (G^{2}_{\mbox{\tiny V}} - G^{2}_{\mbox{\tiny S}})\rho - A\rho^2 - B\rho^3
\label{nlpc-h}
\end{eqnarray}

where the density dependence of the nucleon effective mass $M^*$ reads

\begin{eqnarray}
M^*(\rho)=\frac{M^2}{(M+G^2_{\mbox{\tiny S}}\rho + 2A\rho^2 +3B\rho^3)}.
\end{eqnarray}

In the calculations, we have also used that the scalar density can be approximated by 
$\rho_s=\rho(1-2\tilde{B}_0k^2)$. 

From the single-particle energy in Eq.~(\ref{nlpc-h}), we conclude that the energy of a 
system of $N$ nucleons is 

\begin{equation}
E_N = \frac{2}{M^*}\sum_{i=0}^{k_F}k_i^2 + N[(G^{2}_{\mbox{\tiny V}}-G^{2}_{\mbox{\tiny S}})\rho - A\rho^2 - B\rho^3],
\label{nrl-en}
\end{equation}

where $k_F$ is the Fermi momentum and, due to the Pauli exclusion principle, $4$ 
is the number of nucleons in each energy level. By assuming in one dimension the momentum 
discretization as $k=\frac{2\pi n}{L}$ (periodic conditions), we have 

\begin{eqnarray}
\sum_{i=0}^{k_F}k_i^2=\frac{L}{2\pi}\sum_{i=0}^{k_F} \frac{2\pi}{L}\,k_i^2 = \frac{L}{2\pi}\sum_{i=0}^{k_F}\Delta k\,k_i^2.
\end{eqnarray}

In the continuum limit ($\Delta k\rightarrow 0$) we have

\begin{eqnarray}
\sum_{i=0}^{k_F}k_i^2 \rightarrow \frac{L}{2\pi}\int_{0}^{k_F}k^2\, dk.
\end{eqnarray}

Thus, in three dimensions, 

\begin{eqnarray}
\sum_{i=0}^{k_F}k_i^2 &\rightarrow& \frac{V}{(2\pi)^3}\int k^2\,d^3k 
=\frac{V}{2\pi^2}\frac{k_F^5}{5}
= \frac{3V\lambda}{20}\rho^{\frac{5}{3}},\quad
\end{eqnarray}

where $\lambda=(3\pi^2/2)^{\frac{2}{3}}$ and $V=L^3$ is the system volume. By 
applying such analysis to Eq.~(\ref{nrl-en}), we can finally write the system energy 
density, $\varepsilon=E_N/V$, as

\begin{eqnarray}
\varepsilon^{\mbox{\tiny (NR)}}&=& 
\frac{3\lambda}{10M^*}\rho^{\frac{5}{3}} + (G^{2}_{\mbox{\tiny V}}-G^{2}_{\mbox{\tiny S}})\rho^{2} -A\rho^{3} -B\rho^{4}.
\label{nrl-denerg}
\end{eqnarray}

From Eq.~(\ref{nrl-denerg}) it is possible to obtain all remaining thermodynamical 
quantities of the system. For our purposes in this paper, we will focus on the expression 
for the pressure, calculated as $P=\rho^2\frac{\partial(\mathcal{E}/\rho)}{\partial\rho}$. 
Its form is the following,

\begin{eqnarray}
P^{\mbox{\tiny (NR)}}&=& (G^{2}_{\mbox{\tiny V}}-G^{2}_{\mbox{\tiny S}})\rho^{2} 
-2A\rho^{3} -3B\rho^{4}\nonumber \\
&+& \frac{\lambda}{5M^{2}}\left(M +\frac{5}{2}G^{2}_{\mbox{\tiny S}}\rho +8A\rho^{2} 
+\frac{33}{2}B\rho^{3}\right)\rho^{\frac{5}{3}}.\quad
\label{nrl-press}
\end{eqnarray}

For other equations of state derived from the NRL model, including those of the isovector 
sector, such as the one for symmetry energy, its slope and curvature, we address the reader to 
Ref.~\cite{bianca}.

The coupling constants of the model are $G^{2}_{\mbox{\tiny S}}$, $G^{2}_{\mbox{\tiny 
V}}$, $A$, and $B$. They are adjusted in order for the model to present particular 
values of $\rho_o$ (saturation density), $B_o$ (binding energy), $K_o$ (incompressibility) 
and $M^*_o$, with the last three quantities evaluated at $\rho=\rho_o$. This is 
done by solving a system of four equations, namely, \mbox{$\varepsilon^{\mbox{\tiny 
(NR)}}(\rho_o)=-B_o$}, $K^{\mbox{\tiny (NR)}}(\rho_o)=9[\partial 
P^{\mbox{\tiny(NR)}}/\partial\rho]_{\rho_o}=K_o$, $P^{\mbox{\tiny (NR)}}(\rho_o)=0$, and 
$M^*(\rho_o)/M=M^*_o/M\equiv m^*$. Following such a procedure, we are able to construct 
different parametrizations of the NRL model, using as input physical values of the 
observables $\rho_o$, $B_o$, $K_o$ and $m^*$.

\subsection{Finite temperature regime: critical parameters and correlations}

As a first comment on the calculations in the finite temperature regime, we remind the 
reader that in any fermion system with a four-fermion interaction, namely, a contact one 
as in NLPC model or a boson-mediated as in the model we will discuss in the next section, 
there are various zero-sounds in scalar, spin and spin-isospin channels, which do not 
contribute to the ground state at zero temperature, but do so at finite temperatures. For 
the sake of simplicity and as a first approximation, such contributions will be 
disregarded in the present calculations but can be, in principle, important.

\begin{figure}[!htb]
\centering
\includegraphics[scale=0.33]{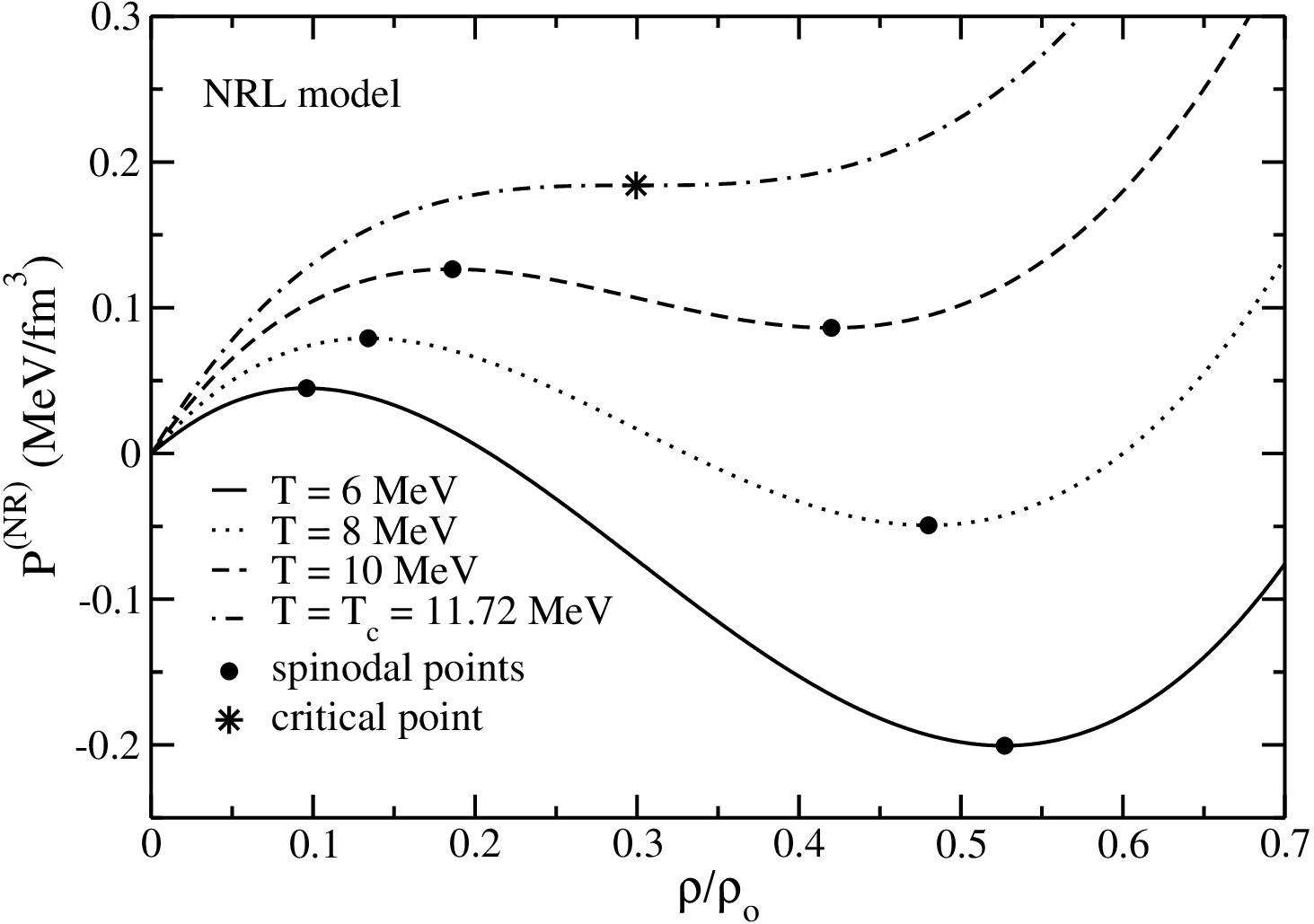}
\vspace{-0.2cm}
\caption{Some isotherms of the NRL model constructed for a parametrization in which 
$\rho_o=0.15$~fm$^{-3}$, $B_o=16$~MeV, $K_o=250$~MeV, and $m^*=0.6$.} 
\label{nlpc-iso}
\end{figure}

In order to investigate possible correlations in the finite temperature regime of the NRL model, we 
proceed to include temperature effects in Eq.~(\ref{nrl-press}) by adding the 
classical ideal gas contribution $\rho T$ as a first approximation, i.e, neglecting any 
quantum fluctuations. This term was inspired by the work of Ref.~\cite{mekjian}. Despite 
this very crude approximation, one can verify from Fig.~\ref{nlpc-iso} that the model 
still presents the qualitative patterns exhibited by hadronic models at finite 
temperatures around $T\lesssim 
20$~MeV~\cite{temp1,temp2,temp3,temp4,temp5,temp6,temp7,rios,temp8,temp9}, i.e., the van 
der Waals-like isotherms at different temperatures with the respective spinodal points 
(points in which $\partial P^{\mbox{\tiny(NR)}}/\partial\rho=0$). 

We also see a critical behavior at a temperature after which the system shows only 
a gaseous nuclear matter phase. This critical temperature, $T=T_c$, characterizes the 
system's critical point (CP), with thermodynamic coordinates $\rho=\rho_c$ and 
$P=P_c$. As another feature, it is worth noticing that all isotherms are confined to a 
region where the densities are always lower than~$\rho_o$, indicating that the liquid-gas 
phase transition occurs always at subsaturation densities, a feature shared by all the usual hadronic 
models. 

In the particular parametrization used in Fig.~\ref{nlpc-iso}, we see that the value 
of the critical temperature lies around $12$~MeV. We highlight that such value also 
depends on the way the equations of state are obtained. In our calculation we are using 
the mean-field approximation. In other approaches, such as the chiral perturbation theory, 
accounting for the inclusion of loop contributions leads to a change of $T_c$ to higher 
values. In Ref.~\cite{fritsch}, for instance, a three-loop calculation of nuclear matter 
produced $T_c=25.5$~MeV.

Still concerning the CP, where $P=P_c$ at the critical density ($\rho_c$) 
and temperature ($T_c$), it also satisfies the condition of vanishing first and second 
derivatives in the $P\times\rho$ function. Therefore, in order to exactly locate the CP, 
it is necessary to impose, simultaneously, the following conditions,

\begin{eqnarray}
P_c=P(\rho_c , T_c),\quad 
\frac{\partial P}{\partial\rho}\bigg|_{\rho_c , T_c}=0,\quad
\frac{\partial^2 P}{\partial\rho^2}\bigg|_{\rho_c , T_c}=0.\quad
\label{conditions}
\end{eqnarray}

For the NRL limit, these conditions lead to the three equation given by

\begin{align}
&2(G^{2}_{\mbox{\tiny V}}-G^{2}_{\mbox{\tiny S}}) 
-12A\rho_c -36B\rho_c^{2} 
\nonumber\\
&+ \frac{2\lambda }{9M^{2}} \left(M +10G^{2}_{\mbox{\tiny S}}\rho_c 
+\frac{352}{5}A\rho_c^{2} 
+\frac{2541}{10}B\rho_c^{3}\right)\rho_c^{-\frac{1}{3}} =0,
\label{sist1}
\end{align}

\begin{align}
T_c &= -2(G^{2}_{\mbox{\tiny V}}-G^{2}_{\mbox{\tiny S}})\rho_c 
+6A\rho_c^{2} +12B\rho_c^{3} 
\nonumber \\
&- \frac{\lambda }{3M^{2}} \left(M +4G^{2}_{\mbox{\tiny S}}\rho_c 
+\frac{88}{5}A\rho_c^{2} 
+\frac{231}{5}B\rho_c^{3}\right)\rho_c^{\frac{2}{3}},
\label{sist2}
\end{align}

and

\begin{align}
P^{\mbox{\tiny (NR)}}_c &= (G^{2}_{\mbox{\tiny V}}-G^{2}_{\mbox{\tiny S}})\rho_c^{2} 
-2A\rho_c^{3} -3B\rho_c^{4} +\rho_c T_c
\nonumber\\
&+ \frac{\lambda }{5M^{2}} \left(M +\frac{5}{2}G^{2}_{\mbox{\tiny S}}\rho_c +8A\rho_c^{2} 
+\frac{33}{2}B\rho_c^{3}\right)\rho_c^{\frac{5}{3}}.
\label{sist3}
\end{align}

\begin{figure}[!htb]
\centering
\includegraphics[scale=0.33]{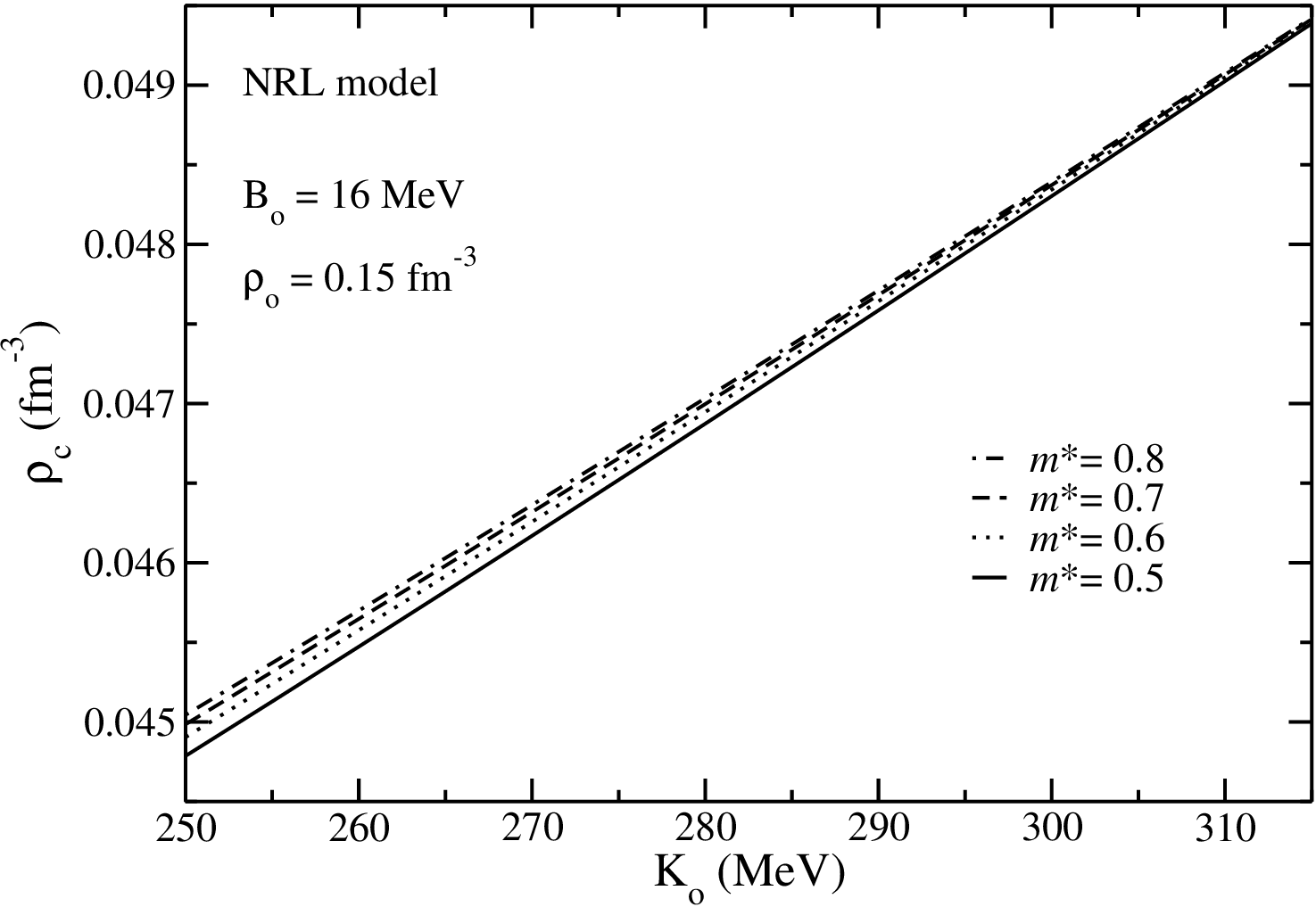}
\vspace{-0.2cm}
\caption{Critical density of the NRL model as a function of incompressibility for 
different effective mass values.} 
\label{nrl-rhoc}
\end{figure}

One can see that, except for $\rho_c$, all critical parameters have an analytical form 
well defined. Thus, for $T_c$ and $P_c$, it is possible to search for functional forms 
relating them to zero temperature bulk quantities. In order to proceed in that direction, 
we need to write the coupling constants of the NRL model, $G^{2}_{\mbox{\tiny S}}$, 
$G^{2}_{\mbox{\tiny V}}$, $A$, $B$, as a function of $\rho_o$, $B_o$, $K_o$ and $m^*$. 
This calculation was already performed in Ref.~\cite{bianca}. It is straightforward 
to implement it in Eqs.~(\ref{sist2}) and (\ref{sist3}). However, we still need to find 
out how $\rho_c$ depends on $\rho_o$, $B_o$, $K_o$ and $m^*$. In order to perform such 
analysis, we first fix the saturation density and binding energy values to those well 
established in the literature, namely, $\rho_o=0.15$~fm$^{-3}$ and $B_o=16$~MeV, to 
specifically search for the function $\rho_c=\rho_c(K_o,m^*)$. Following this route, 
we numerically solve Eq.~(\ref{sist1}) and present in Fig.~\ref{nrl-rhoc} the results of 
$\rho_c$ as a function of $K_o$ for different values of $m^*$.

As shown in Fig.~\ref{nrl-rhoc}, the critical density is much more sensitive to variation of the 
incompressibility than of the effective mass. Furthermore, the $K_o$ variation 
is practically linear. From this result, it is possible to parametrize the $K_o$ 
dependence of $\rho_c$ as follows,

\begin{eqnarray}
\rho_c = \alpha + \beta K_o, 
\label{nrlrhoc}
\end{eqnarray}

with $\alpha=(0.0278\pm 1.34\times 10^{-4})$~fm$^{-3}$, and $\beta=(6.84\times 10^{-5}\pm 
4.76\times 10^{-7})$~MeV$^{-1}$~fm$^{-3}$. Thus, the use of this $\rho_c(K_o)$ function 
in Eq.~(\ref{sist2}), along with the expressions of $G^{2}_{\mbox{\tiny S}}$, 
$G^{2}_{\mbox{\tiny V}}$, $A$ and $B$ as a function of $\rho_o$, $B_o$, $K_o$ and $m^*$, 
leads to the following analytical expressions for $T_c$,

\begin{align}
T_c &= a_0(\alpha +\beta K_o)^\frac{2}{3}
+ \sum_{n=1}^{6}\frac{a_n}{t}(\alpha + \beta K_o)^m (t_{n1}K_o + t_{n2})\nonumber\\
&-\frac{1}{m^*}\sum_{n=1}^{6}\frac{b_0}{t}b_n(\alpha +\beta K_o)^m.
\label{nrl-tc}
\end{align}

In this expression, $m=n$ for $n\leqslant 3$, and $m=n-7/3$ for $n>3$. One also has 
that $t = 3M^2 - 19E_{\mbox{\tiny F}}^oM + 18E_{\mbox{\tiny F}}^{o2}$, with 
$E_{\mbox{\tiny F}}^o=3\lambda\rho_{o}^{\frac{2}{3}}/10M$. The coefficients are listed in 
the Appendix. It is worth to notice that in order for $T_c$ to be given in MeV, we need to 
convert $\alpha$ and $\beta$ to appropriate units. Such a conversion leads to 
$\alpha=(59.8\pm10.1)^3$~MeV$^3$ and $\beta=(22.9\pm 1.91)^2$~MeV$^2$. In these units, 
the densities are given in MeV$^3$.

\begin{figure}[!htb]
\centering
\includegraphics[scale=0.33]{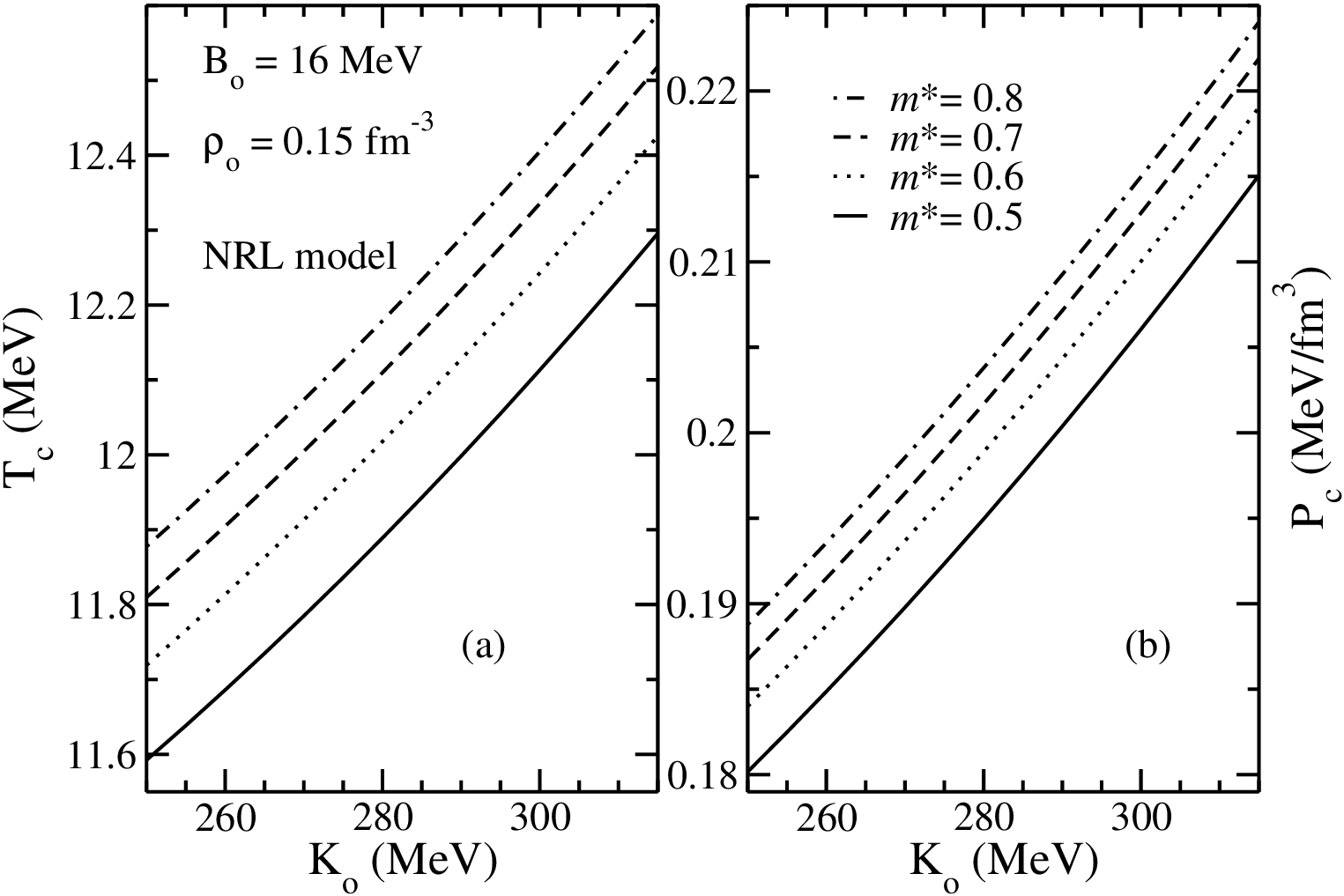}
\vspace{-0.2cm}
\caption{Critical (a) temperature, and (b) pressure of the NRL model as a function of 
incompressibility for different effective mass values.} 
\label{nrl-tcpc}
\end{figure}

Following the same procedure in Eq.~(\ref{sist3}), we also found an analytical form for 
the critical pressure in the NRL model. The result is,

\begin{align}
P_c &= c_0(\alpha +\beta K_o)^\frac{5}{3}
+ \sum_{n=1}^{6}\frac{c_n}{t}(\alpha + \beta K_o)^l (t_{n1}K_o + t_{n2})\nonumber\\
&-\frac{1}{m^*}\sum_{n=1}^{6}\frac{d_0}{t}d_n(\alpha +\beta K_o)^l,
\label{nrl-pc}
\end{align}

where $l=n+1$ for $n\leqslant 3$, and $l=n-4/3$ for $n>3$. For complete expressions of the 
coefficients, including its $\rho_o$ and $B_o$ dependence, we address the reader to 
the Appendix.
The incompressibility dependence of $T_c$ and $P_c$ is displayed in Fig.~\ref{nrl-tcpc} 
for some fixed values of $m^*$.

As we see in Fig.~\ref{nrl-tcpc}, $T_c$ and $P_c$, as well as 
$\rho_c$ are increasing functions of the incompressibility. On the 
other hand, critical temperature and pressure are more sensitive to effective mass effects
than the critical density. Furthermore, $T_c$ and $P_c$ are also increasing functions of 
$m^*$. In next section, we will verify if such patterns are also exhibited in RMF models.

\section{RMF parametrizations analysis} 

\label{rmf-analysis}

\subsection{Theoretical framework}

In the original Walecka model~\cite{walecka}, there are two free parameters adjusted to 
impose the values 
of two particular observables of infinite nuclear matter, namely, $\rho_o$ ($\sim 
0.15$~fm$^{-3}$) and $B_o$ ($\sim16$~MeV). However, the model fails in the description of the 
incompressibility and effective mass ratio ($m^*=M^*/M$) at the saturation density, for the results 
for their values lie close to $550$~MeV and $0.54$, respectively. In order to solve that problem, 
Boguta and Bodmer~\cite{boguta} have introduced in the original Walecka model two 
additional terms 
representing cubic and quartic self-interactions in the $\sigma$ field, providing two new free 
parameters now adjusted to correctly reproduce $K_o$ and $m^*$. The Lagrangian density of the 
Boguta-Bodmer (BB) model is,

\begin{align}
\mathcal{L} &= \overline{\psi}(i\gamma^\mu\partial_\mu - M)\psi + 
g_\sigma\sigma\overline{\psi}\psi + \frac{1}{2}(\partial^\mu \sigma \partial_\mu \sigma - 
m^2_\sigma\sigma^2) 
\nonumber \\
&- \frac{\mathcal{A}}{3}\sigma^3 - \frac{\mathcal{B}}{4}\sigma^4 
- g_\omega\overline{\psi}\gamma^\mu\omega_\mu\psi -\frac{1}{4}F^{\mu\nu }F_{\mu\nu} + 
\frac{1}{2}m^2_\omega\omega_\mu\omega^\mu,
\label{rmf-lag}
\end{align}

with $F_{\mu\nu}=\partial_\nu\omega_\mu-\partial_\mu\omega_\nu$. The free parameters are 
$g_\sigma$, $g_\omega$, $g_\rho$, $\mathcal{A}$ and $\mathcal{B}$.

Since the original work of Boguta and Bodmer~\cite{boguta} published in 1977, many parametrizations 
of the BB model were proposed along the years. For a list of $128$ of them, collected in 
a unique reference, we address the reader to Ref.~\cite{rmf}. In the notation of that 
paper, the authors named the BB parametrization as {\it type 2} ones. Those obtained from 
the original Walecka model are called {\it type 1} parametrizations.

From Eq.~(\ref{rmf-lag}), it is possible to construct all thermodynamical quantities at 
zero and finite temperature by following, for instance, the steps shown in Ref.~\cite{serot}. For 
our purposes in this paper, we only show the pressure of symmetric ($\gamma=4$) 
infinite nuclear matter, that reads,

\begin{align}
P &= \dfrac{G_\omega^2\rho^2}{2} - \dfrac{(\Delta M)^2}{2G_\sigma^2} 
+ \dfrac{g_3(\Delta M)^3}{3} - \dfrac{g_4(\Delta M)^4}{4} \nonumber \\
&+ \dfrac{\gamma}{6\pi^2}\int_0^{\infty}\dfrac{dk\,k^4}{(k^2 + 
{M^*}^2)^{1/2}}\left[n(k,T,\nu)+\bar{n}(k,T,\nu)\right].
\label{rmf-press}
\end{align}

with $\Delta M=M^*-M$. The Fermi-Dirac distributions for particles and antiparticles are, 
respectively,

\begin{align}
n(k,T,\nu) &= \frac{1}{e^{(E^*-\nu)/T}+1}\quad\mbox{and}\nonumber\\
\bar{n}(k,T,\nu) &= \frac{1}{e^{(E^*+\nu)/T}+1},
\end{align}

with $E^*=(k^2+{M^*}^2)^{1/2}$. The effective mass and chemical potential are given by

\begin{align}
M^* &= M + g_\sigma \langle\sigma\rangle \nonumber\\
&= M - G_\sigma^2\left[\rho_s - g_3(\Delta M)^2 + g_4(\Delta M)^3\right],
\label{rmf-effmass}
\end{align}

and $\nu=\mu-G_\omega^2\rho$. The vector and scalar densities are also written in terms 
of $n$ and $\bar{n}$ as follows,

\begin{align}
\rho&=\dfrac{\gamma}{2\pi^2}\int_0^{\infty}dk\,k^2
\left[n(k,T,\nu)-\bar{n}(k,T,\nu)\right],
\label{rmf-rho}
\nonumber \\
\rho_s &= \dfrac{\gamma}{2\pi^2}\int_0^{\infty}\frac{dk\,M^*k^2}{(k^2+{M^*}^2)^{1/2}}
\left[n(k,T,\nu)+\bar{n}(k,T,\nu)\right].
\end{align}

Finally, the new free parameters present in Eqs.~(\ref{rmf-press}), 
(\ref{rmf-effmass}), and in definition of $\nu$, are defined in terms of the previous ones 
as $G_\sigma^2 = \frac{g_\sigma^2}{m_\sigma^2}$, $G_\omega^2 = 
\frac{g_\omega^2}{m_\omega^2}$, $g_3 = \frac{\mathcal{A}}{g_\sigma^3}$, and $g_4 = 
\frac{\mathcal{B}}{g_\sigma^4}$.

\subsection{Correlation of critical parameters}

\begin{figure}[!htb]
\centering
\includegraphics[scale=0.35]{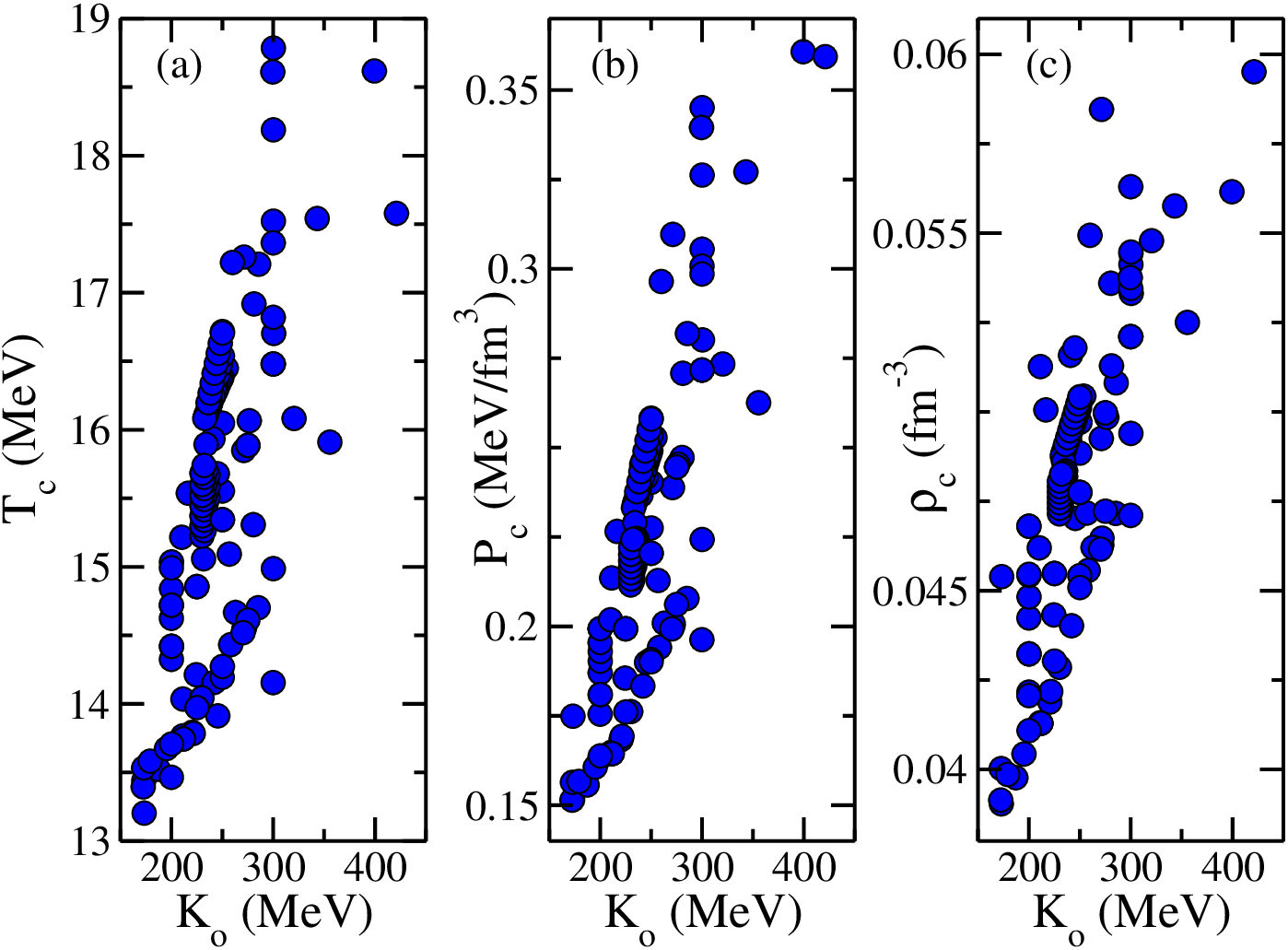}
\vspace{-0.2cm}
\caption{Critical (a) temperature, (b) pressure, and (c) density of the RMF BB 
parametrizations collected in Ref.~\cite{rmf}.} 
\label{rmf-critical}

\end{figure}

We are now able to search for possible correlations between critical 
parameters of BB parametrizations. As a starting point, we remark that in 
Ref.~\cite{bianca}, our results indicate correlations between zero temperature bulk 
parameters in the NRL model that are also reproduced specifically in the parametrizations 
of the BB model. As an example, in that paper we found, for the isovector sector of 
the NRL model, that $L_o$ (symmetry energy slope at $\rho_o$) is linearly correlated with 
$J$ (symmetry energy at $\rho_o$) for those parametrizations presenting fixed values of 
$m^*$ and $K_o$. We also found the same correlation conditions for the BB model. Many 
other bulk parameters, including those from the isoscalar sector, present such a pattern 
concerning correlations of the BB model and its nonrelativistic version (the NRL model). 
In that sense, we have used the NRL model as a guide to investigate correlations 
in the BB model. Here we proceed in the same direction but now regarding correlations 
between finite and zero temperature quantities. Based on this discussion and applying 
the critical condition of Eq.~(\ref{conditions}), we calculated the critical parameters of 
the $128$ BB parametrizations of Ref.~\cite{rmf}, in order to see some evidence of 
correlations. The results are shown in Fig.~\ref{rmf-critical}.

\begin{figure}[!htb]
\centering
\includegraphics[scale=0.35]{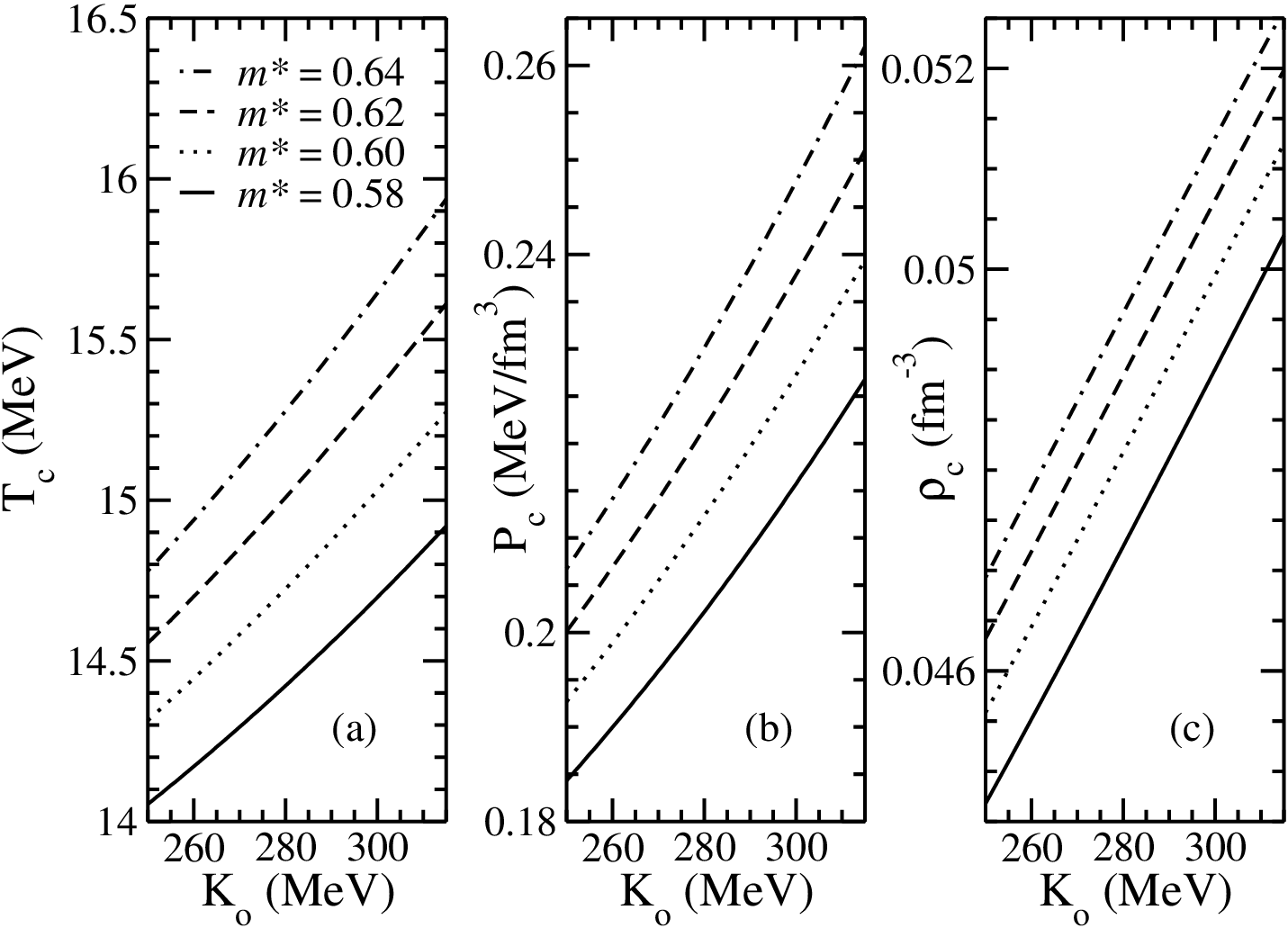}
\vspace{-0.2cm}
\caption{Critical parameters of BB parametrizations in which $\rho_o=0.15$~fm$^{-3}$ and 
$B_o=16$~MeV, for fixed values of $m^*$.} 
\label{rmf-corr-Ko}
\end{figure}

We see that the critical parameters seem to indicate an increasing trend as $K_o$ 
increases. However, the almost linear pattern exhibited in the NRL parametrizations, or 
more precisely, a clear connection with $K_o$, is not observed, as a simple comparison 
between Figs.~\ref{nrl-rhoc} and~\ref{nrl-tcpc} suggests. Therefore, we proceed to impose 
the condition of fixed values for $m^*$ as we did in the NRL case. In order to perform 
such analysis, we construct BB parametrizations in which $\rho_o=0.15$~fm$^{-3}$, 
$B_o=16$~MeV, and for the two remaining observables, namely, $m^*$ and $K_o$, we 
investigate models in particular ranges. Actually, here we adopt the same constraints 
used 
in Ref.~\cite{bianca}, i.e., for the effective mass ratio, $0.58\leqslant m^*\leqslant 
0.64$, and for the incompressibility, $250\leqslant K_o \leqslant 315$~MeV. According to 
Ref.~\cite{ls-splitting}, the former constraint allows parametrizations of the BB model to 
present spin-orbit splittings in agreement with well-established experimental values for 
$^{16}\rm{O}$, $^{40}\rm{Ca}$, and $^{208}\rm{Pb}$ nuclei. The latter constraint, on the 
other hand, was generated in a recent study~\cite{stone} where the authors based their 
calculations on a reanalysis of up-to-date data on isoscalar giant monopole resonance 
energies of Sn and Cd isotopes. They claimed that such a range, close to the $K_o$ value 
of many RMF parametrizations, was obtained without any microscopic assumptions and is 
basically due to the suitable treatment of nuclear surface properties. Based on this 
discussion we show the critical parameters of BB parametrizations in 
Fig.~\ref{rmf-corr-Ko}.

\begin{figure}[!htb]
\centering
\includegraphics[scale=0.348]{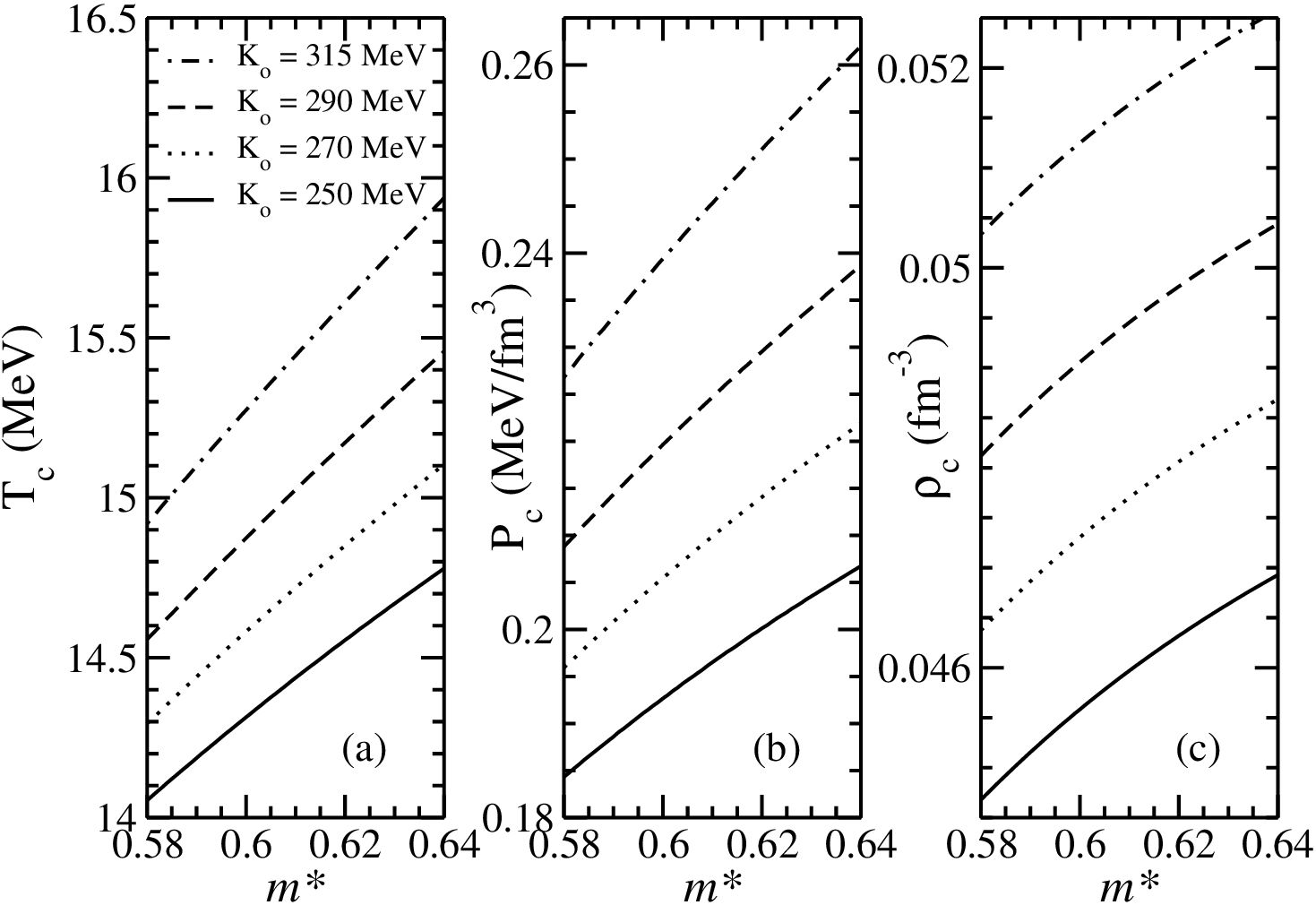}
\vspace{-0.2cm}
\caption{Critical parameters of BB parametrizations in which $\rho_o=0.15$~fm$^{-3}$ and 
$B_o=16$~MeV, for fixed values of $K_o$.} 
\label{rmf-corr-m}
\end{figure}

We see in Fig.~\ref{rmf-corr-Ko}{\color{blue}(a)} that the $K_o$ dependence of $T_c$ is 
qualitatively the same as in the NRL model, see Fig.~\ref{nrl-tcpc}{\color{blue}(a)}. The 
correlation between these quantities is verified for fixed values of the effective mass, with 
$T_c$ being an increasing function of $K_o$. The same pattern is also observed for both $P_c$ 
and $\rho_c$, as seen in Figs.~\ref{rmf-corr-Ko}{\color{blue}(b)} and 
\ref{rmf-corr-Ko}{\color{blue}(c)}, respectively. The behavior of these latter critical 
parameters was also pointed out by the NRL model, as shown in 
Figs.~\ref{nrl-tcpc}{\color{blue}(b)} and \ref{nrl-rhoc}. For the sake of completeness, we 
also display in Fig.~\ref{rmf-corr-m} the effective mass dependence of the critical 
parameters. 

It is verified that they are also tightly correlated. By comparing these results with those 
from Sec.~\ref{nrl-analysis}, we see that $T_c$ and $P_c$ of the NRL limit model also 
depend on~$m^*$, as in the relativistic case, but $\rho_c$ is practically not affected, 
see Fig.~\ref{nrl-rhoc}. The source of such a difference might be attributed to the fact 
that, in the NRL model at finite temperature, we did not take into account the $T$ 
dependence of $\rho$, like in the relativistic case, see Eq.~(\ref{rmf-rho}). If we had 
done so, we would have $\rho=\rho(E_{kin},T)$ with the kinetic energy $E_{kin}$ being 
a function of the effective mass, see the first term of Eq.~(\ref{nlpc-h}). Thus, in the 
NRL model, the effect of $m^*$ on $\rho_c$ is underestimated in comparison with the 
relativistic case.

Based on these results, one can see that most of the correlations present in the NRL model at 
finite temperature regime, namely, critical parameters related to saturation bulk 
quantities at zero temperature, are reproduced in the parametrizations of the BB model, 
since one preserves the same conditions that drive the arising of such correlations. 
These conditions explain why we do not see a tight correlation of $T_c$, $P_c$ and 
$\rho_c$ with $K_o$, for instance, in the BB parametrizations of Fig.~\ref{rmf-critical}. 
In that case, besides having different values of $\rho_o$ and $B_o$, each parametrization 
presents a particular value of effective mass, and do not satisfy the condition of fixed 
$m^*$, constraint that produces a clear connection between the critical parameters 
and~$K_o$. Similar analysis can be performed in order to describe the correlation of the 
critical parameters and $m^*$. In this case, it is established if the condition of having 
BB parametrizations presenting the same value of $K_o$ is fulfilled.

Still regarding our findings on the correlations presented here, and in order to clarify 
our discussion, we remind the reader that they were found for the specific RMF model 
presenting the self-coupling in the scalar field up to fourth order. We are dealing 
with parametrizations of the BB model in which the equations of state were obtained 
through the widely used mean-field approximation (MFA). Therefore, it is not our purpose 
to classify them as universal. A more detailed study based on other kind of models 
described by more sophisticated Lagrangian densities in comparison with that of 
Eq.~(\ref{rmf-lag}) is in order. Even calculations that go beyond MFA can change the 
correlations found here, stressing the importance of performing such an investigation in order to 
establish possible correlations between zero and finite temperature quantities in different kind of 
hadronic models.

\subsection{Comparison with other theoretical studies}

Specifically concerning the relation between $T_c$ and $K_o$, we remark here 
that our findings for the RMF parametrizations analyzed here are in qualitative agreement 
with previous studies on such correlation, as we will show in the following. In the 
Kapusta model of Ref.~\cite{kapusta}, for instance, the author derived an expression for 
the pressure, based on the Sommerfeld expansion in the degenerate regime (Fermi energy 
$\gg$ temperature), that reads $P = K_o\rho^2(\rho-\rho_o)/9\rho_o^2 + 
b^2M^*_o\rho^{1/3}T^2/6$, with $b=1.809$. This leads to a critical temperature of 

\begin{eqnarray}
T_c^{\rm K}=0.326\rho_o^{1/3}\sqrt{K_o/M^*_o},
\label{tck}
\end{eqnarray}

with $T_c$ being an increasing function of $K_o$. In Ref.~\cite{lattimer}, 
Lattimer and Swesty modified the Kapusta expression for the critical temperature by 
introducing an opposite dependence of the saturation density, but keeping the increasing 
pattern concerning $K_o$. The correlation reads 

\begin{eqnarray}
T_c^{\rm LS} = C\rho_o^{-1/3}\sqrt{K_o},
\label{tcls}
\end{eqnarray}

where $C=0.608$~MeV$^{1/2}$~fm$^{-1}$. In another study, Natowitz et al.~\cite{natowitz} 
proposed the inclusion of effective mass effects on the latter correlation, which 
produced the expression 

\begin{eqnarray}
T_c^{\rm N} = C'\rho_o^{-1/3}\sqrt{K_o/m^*},
\label{tcn}
\end{eqnarray}

with $C' = 0.484\pm0.074$~MeV$^{1/2}$~fm$^{-1}$. Finally in Ref.~\cite{rios}, Rios 
improved 
the Kapusta model by introducing, in the pressure equation of state, the density dependence 
of $M^*$ coming from the Skyrme interaction. The result of such improvement generated the 
following correlation, 

\begin{eqnarray}
T_c^{\rm R} = 0.326\bar{m}^*\rho_o^{1/3}\sqrt{K_o/M},
\label{tcr}
\end{eqnarray}

where $\bar{m}^* = m^*(\rho=\frac{5\rho_o}{12})$. Again, we have here the by now familiar pattern of an increasing 
$T_c$ as a function of~$K_o$. It is then fair to say that the BB parametrizations share with former hadronic models the 
qualitative prediction of an increasing $T_c$ as $K_o$ increases.

Regarding the $T_c\times m^*$ correlation, the BB parametrizations also 
show an increasing pattern for $T_c$ (and also for $P_c$ and $\rho_c$). However, the only 
prediction compatible with such behavior is that from Ref.~\cite{rios}. In that case, 
$T_c^{\rm R}$ increases as $\bar{m}^*$ increases, but, for this particular analysis, the 
correlation is with the effective mass ratio evaluated at a subsaturation density of 
$\rho=5\rho_o/12$, and not exactly at $\rho=\rho_o$ as in the case presented in our 
work.

\begin{figure}[!htb]
\centering
\includegraphics[scale=0.35]{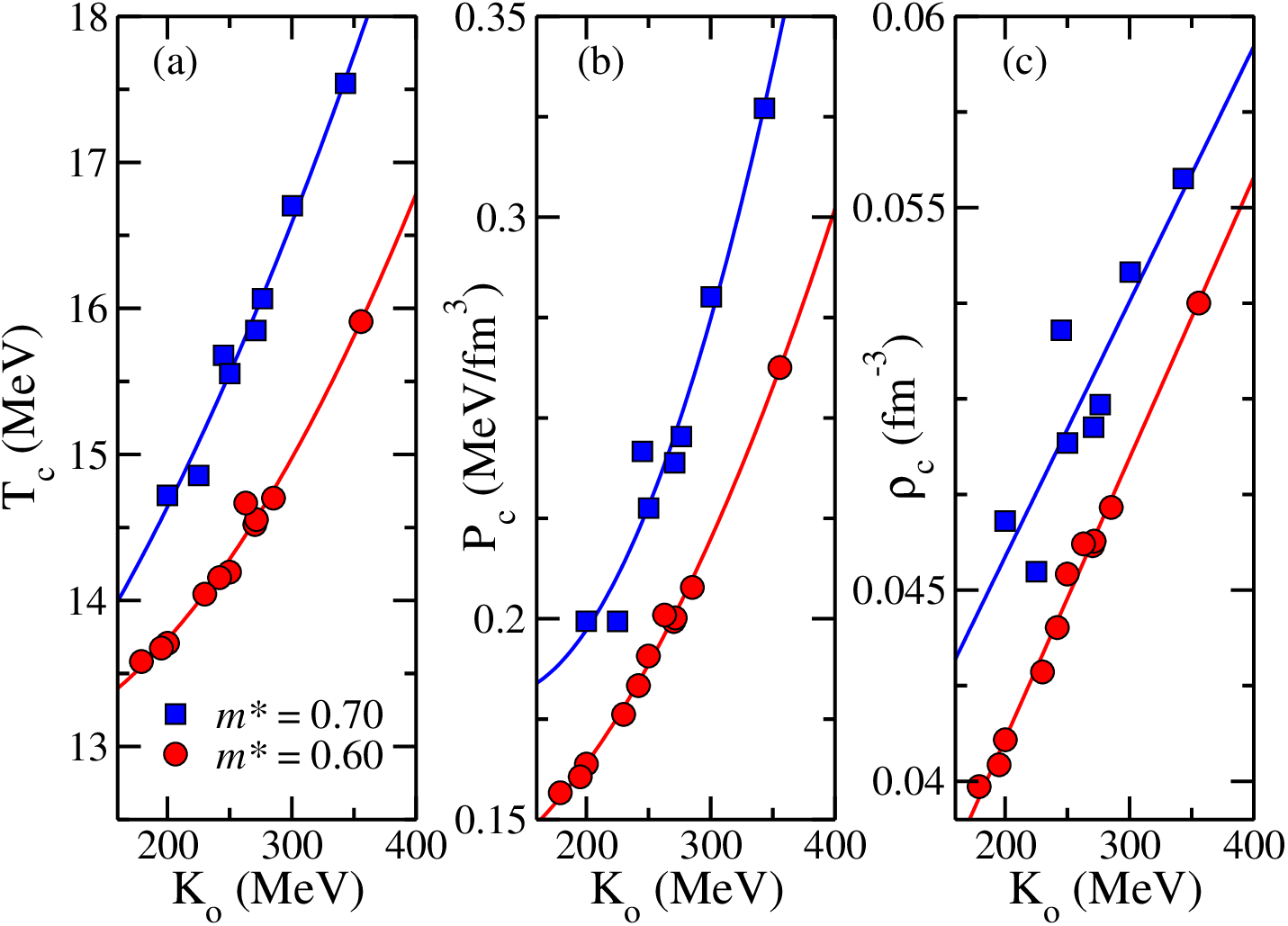}
\vspace{-0.2cm}
\caption{Critical parameters as a function of $K_o$ of some BB parametrizations of 
Ref.~\cite{rmf} presenting $m^*=0.60$ (MS2, NL4, NLSH, NLRA1, Q1, NL3, Hybrid, FAMA1, 
NL-VT1, NL06, NLS) and $m^*=0.70$ (S271, P-070, NLM6, NLD, NL07, GM1, GL4, FAMC2). Full 
lines: fitting curves.} 
\label{rmf-corr-Ko2}
\end{figure}

We have also verified the effect of $K_o$ on the critical parameters of the BB model in some 
known (and largely used) parametrizations of Ref.~\cite{rmf}. The results are depicted in 
Fig.~\ref{rmf-corr-Ko2}. 

\begin{figure}[!htb]
\centering
\includegraphics[scale=0.34]{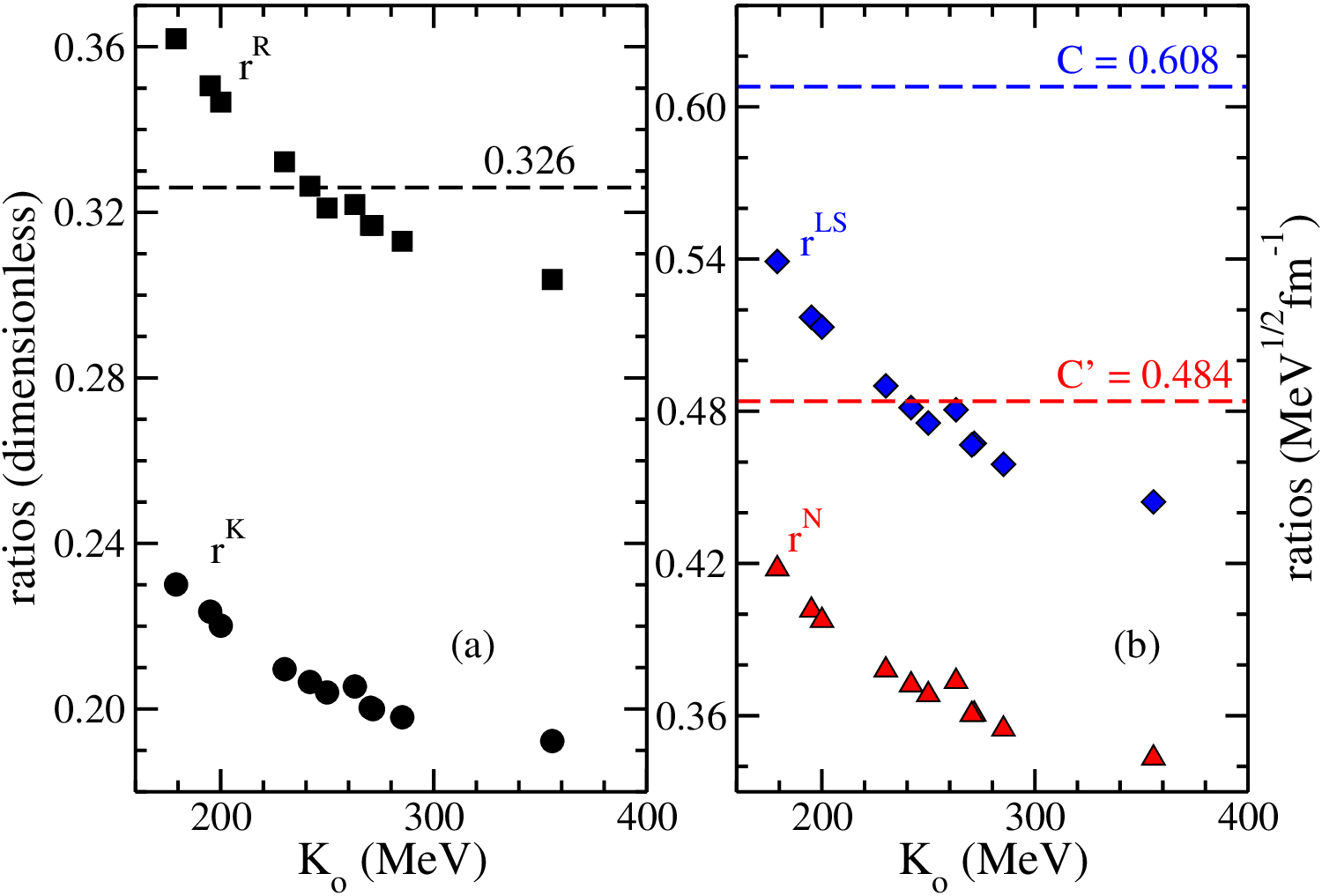}
\vspace{-0.2cm}
\caption{Ratios defined in Eqs.~(\ref{rk})-(\ref{rr}) in comparison with the respective 
constants presented in Eqs.~(\ref{tck})-(\ref{tcr}) for the BB parametrizations of 
Fig.~\ref{rmf-corr-Ko2} in which $m^*=0.60$.} 
\label{rmf-ratios}
\end{figure}

As we can see from this figure, the BB parametrizations present a faster increasing 
of $T_c$ with an increase of $K_o$ in comparison with the previous investigations showed 
by Eqs.~(\ref{tck})-(\ref{tcr}). Our results indicate \mbox{$T_c\sim K_o^a$}, with 
$a\geqslant 1$, different from that pointed out in the aforementioned expressions, namely, 
\mbox{$T_c\sim K_o^{1/2}$}. For those parametrizations in which $m^*=0.60$, for instance, 
we found a fitting curve of \mbox{$T_c = 13 - (1.9\times10^{-3})K_o + 
(2.9\times10^{-5})K_o^2$}. In order to become clearer this difference, we plot in 
Fig.~\ref{rmf-ratios} the following ratios

\begin{eqnarray}
r^{\rm K} &=& \frac{T_c}{\rho_0^{1/3}\sqrt{K_0/M_0^*}}, 
\label{rk} \\
r^{\rm LS} &=& \frac{T_c}{\rho_0^{-1/3}\sqrt{K_0}}, \\ 
r^{\rm N} &=& \frac{T_c}{\rho_0^{-1/3}\sqrt{K_0/m^*}}, \quad \mbox{and} \\
r^{\rm R} &=& \frac{T_c}{\bar{m}^*\rho_0^{1/3}\sqrt{K_0/M}}.
\label{rr}
\end{eqnarray}

The comparison of these ratios with the ones derived from 
Eqs.~(\ref{tck})-(\ref{tcr}) shows explicitly the deviations between these different 
approaches.

By returning to Fig.~\ref{rmf-corr-Ko2}, one notice also some deviations from the 
fitting curves for those parametrizations presenting $m^*=0.70$. We attribute such 
differences to the distinct values of $\rho_o$ and $B_o$ presented by each model. For 
those in which $m^*=0.70$, the variation of $\rho_o$ and $B_o$ is larger than those 
presenting $m^*=0.60$. The former has $\Delta\rho_o=0.015$~fm$^{-3}$ and $\Delta 
B_o=0.72$~MeV, and the latter, $\Delta\rho_o=0.004$~fm$^{-3}$ and $\Delta B_o=0.69$~MeV. 
In the BB parametrizations analyzed in Figs.~\ref{rmf-corr-Ko} and~\ref{rmf-corr-m}, we 
did not see any deviation due to the fact that we have fixed the values of saturation density 
and binding energy, i. e., we had $\Delta\rho_o=\Delta B_o=0$ in all cases. The effects 
induced specifically by the variations of $\rho_o$ and $B_o$ in the critical parameters 
can be seen in the next Figs.~\ref{rmf-corr-rho} and~\ref{rmf-corr-b}.

\begin{figure}[!htb]
\centering
\includegraphics[scale=0.347]{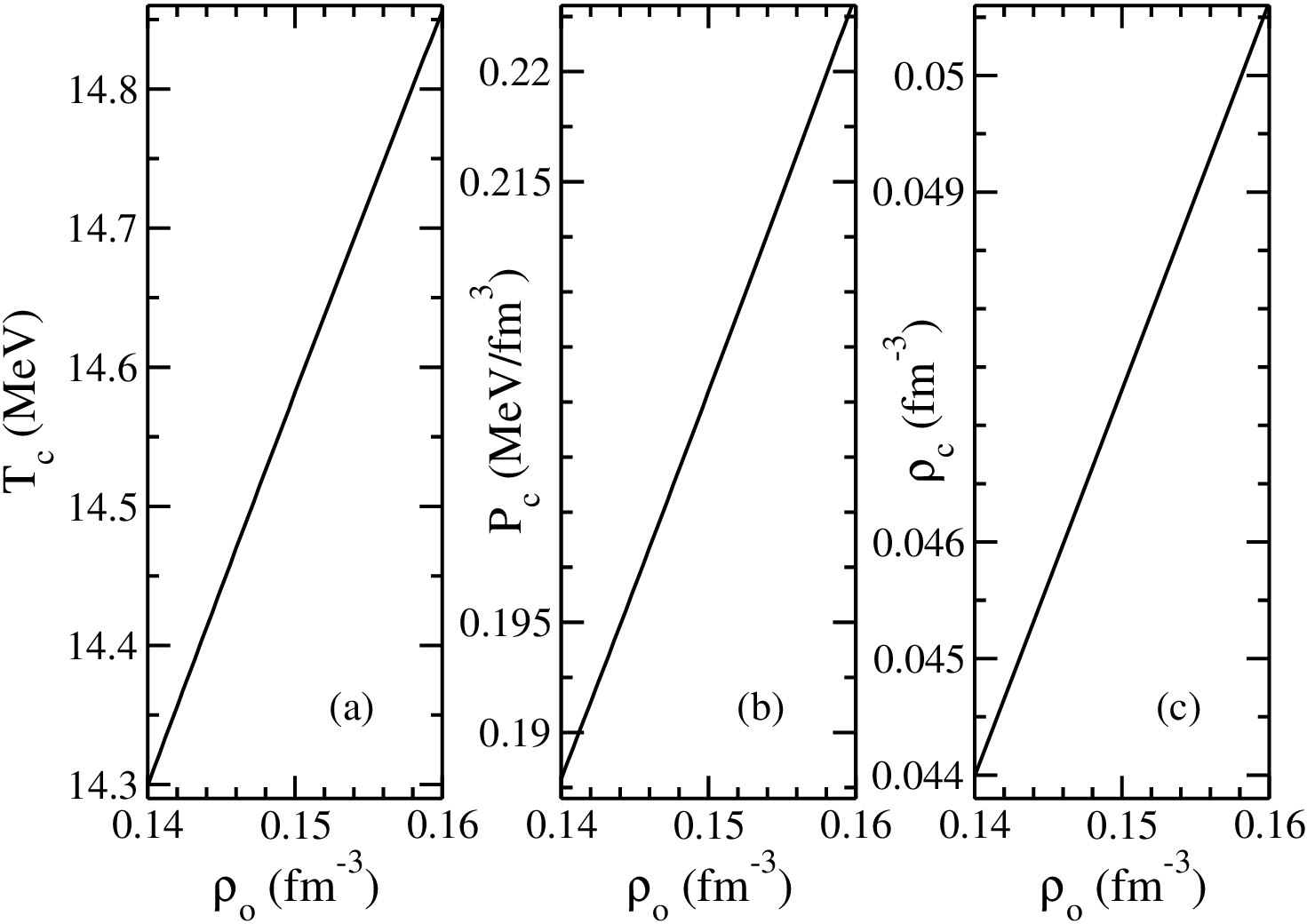}
\vspace{-0.2cm}
\caption{Critical parameters as a function of $\rho_o$ for $B_o = 16$~MeV, $m^*=0.6$ 
and $K_o=270$~MeV.} 
\label{rmf-corr-rho}
\end{figure}

\begin{figure}[!htb]
\centering
\includegraphics[scale=0.347]{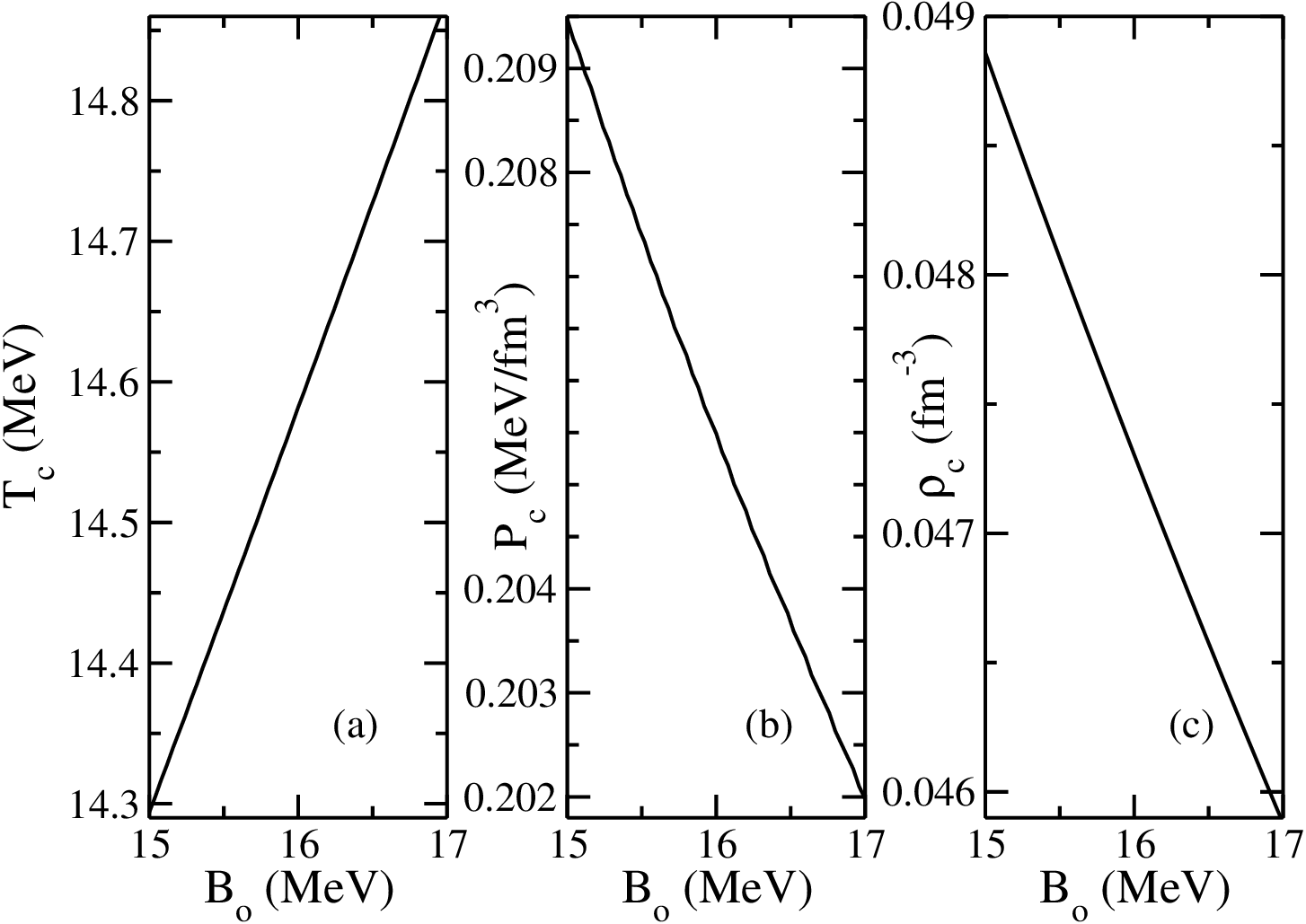}
\vspace{-0.2cm}
\caption{Critical parameters as a function of $B_o$ for $\rho_o = 0.15$~fm$^{-3}$, 
$m^*=0.6$ and $K_o=270$~MeV.} 
\label{rmf-corr-b}
\end{figure}

In Fig.~\ref{rmf-corr-rho}, we see an increasing effect of $\rho_o$ in the critical 
parameters with a linear dependence in all three quantities. The pattern observed in the critical 
temperature specifically is also observed in the correlation found in the 
Kapusta~\cite{kapusta} and Rios~\cite{rios} models, although they have obtained an 
analytical form of $\rho_o^{1/3}$ that differs from the result of the BB parametrizations. 
In the Lattimer-Swesty and Natowitz models, on the other hand, an opposite effect is found, 
since now $T_c$ is proportional to $\rho_o^{-1/3}$, i. e., a decreasing function of 
saturation density. 

In Fig.~\ref{rmf-corr-b}, the increasing pattern is obtained only for $T_c$. The two 
other critical parameters are decreasing functions of $B_o$. Furthermore, for the 
values used in Figs.~\ref{rmf-corr-rho} and~\ref{rmf-corr-b}, we see that $P_c$ and 
$\rho_c$ are less sensitive to the variation of $B_o$ than of $\rho_o$. For a range of 
around $6\%$ in the central value of $B_o=16$~MeV, the changes found in critical pressure 
and density are $\Delta P_c=0.007$~MeV/fm$^3$ and $\Delta\rho_c=0.003$~fm$^{-3}$, 
respectively, while a range of around $7\%$ in $\rho_o=0.15$~fm$^{-3}$ produces $\Delta 
P_c=0.035$~MeV/fm$^3$ and $\Delta\rho_c=0.007$~fm$^{-3}$, i. e., about five and two times higher 
variations, respectively. Regarding the critical temperature, $\Delta T_c$ is practically the same 
for the two cases.

\subsection{Comparison with experimental data}

\begin{figure}[!htb]
\centering
\includegraphics[scale=0.35]{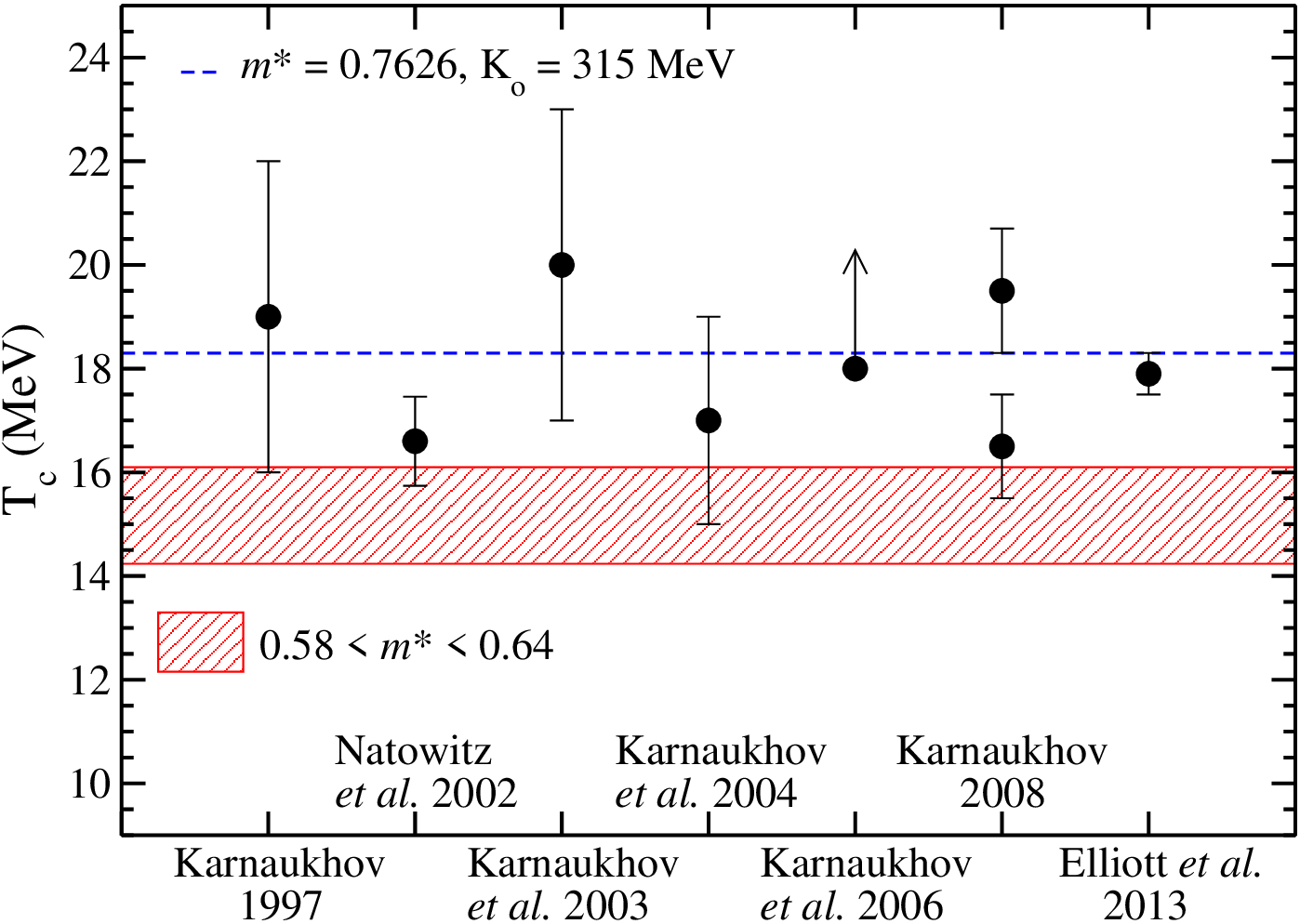}
\vspace{-0.2cm}
\caption{Theoretical predictions (band and dashed line, for $\rho_o=0.153$~fm$^{-3}$ and 
$B_o=16.32$~MeV) and experimental data (circles) on critical temperature of symmetric 
nuclear matter. The references are 
\mbox{Karnaukhov 1997: \cite{karn1}}, 
\mbox{Natowitz {\it et al.} 2002: \cite{natowitz}}, 
\mbox{Karnaukhov {\it et al.} 2003: \cite{karn2}},
\mbox{Karnaukhov {\it et al.} 2004: \cite{karn3}},
\mbox{Karnaukhov {\it et al.} 2006: \cite{karn4}},
\mbox{Karnaukhov 2008: \cite{karn5}}, and
\mbox{Elliott {\it et al.} 2013: \cite{elliott}}.
} 
\label{tc-exp}
\end{figure}

For a direct application of our findings, we use the correlations exhibited by the BB
model to generate parametrizations in which critical parameters can be compared with 
experimental data reported in different works along the years. Specifically regarding the 
critical temperature, many studies have been successful in obtaining this 
quantity. In summary, a beam of relativistic incident light particles that transfers excitation 
thermal energy~($E^*$) is used in order to heat a nucleus. 
The relationship between $E^*$ and $T$ is found through the so called caloric 
curve. This heating procedure gives rise to different emission processes, namely; gamma 
rays emission, occurring for $1 \lesssim T \lesssim2$~MeV; nucleon-evaporation, in the 
range of $2 \lesssim T \lesssim 5$~MeV; and multifragmentation, for $T\gtrsim 5$~MeV, 
this latter process being one that generates emission of $\alpha$~particles, nucleons, and 
intermediate mass fragments (IMF). Theoretical models are commonly used to fit 
experimental data of IMF charge distributions by having the critical temperature as a 
free parameter. Thus, the value of $T_c$ is indirectly calculated. In Fig.~\ref{tc-exp}, 
we compare our theoretical predictions with experimental data obtained for~$T_c$.

In this figure, the horizontal band bounds the possible values of $T_c$ for BB parametrizations 
presenting effective mass in the range of $0.58\leqslant m^*\leqslant 
0.64$~\cite{ls-splitting}, and incompressibility within $250\leqslant K_o \leqslant 
315$~MeV~\cite{stone}. One can see that such parametrizations are compatible with four 
experimental points, by taking into account the error bars. However, if we simply discard 
the constraint related to $m^*$, it is possible to construct a BB parametrization in 
which six (of seven) experimental points are reproduced, including the more recent data on 
the critical parameters obtained in Ref.~\cite{elliott} where $T_c=17.9\pm 0.4$~MeV. In 
Fig.~\ref{tc-exp}, we indicate by the dashed blue line this specific parametrization presenting $m^* 
= 0.7626$ and $K_o=315$~MeV. We see that this choice is more compatible with the trend of higher 
values for $T_c$ pointed out by the experimental data.

Finally, concerning critical pressure and density, we have also verified that the constraints on 
$m^*$ and $K_o$ produce BB parametrizations presenting the range of $0.19\leqslant P_c \leqslant 
0.27$~MeV/fm$^3$ and the value of $\rho_c=0.05$~fm$^{-3}$ (with one significant figure), 
respectively. By comparing such values with Ref.~\cite{elliott}, that found ranges of 
$P_c=0.31\pm0.07$~MeV/fm$^3$ and $\rho_c=0.06\pm0.01$~fm$^{-3}$ from experimental analysis of 
compound nuclear and multifragmentation reactions, we found an overlap of around $21\%$ with former 
range, and agreement within the error bar with the latter one. Moreover, for the parametrization
represented in Fig.~\ref{tc-exp} by the dashed line, where the effective mass constraint is 
neglected, we found critical pressure and density given, respectively, by 
$P_c=0.34$~MeV/fm$^3$ and $\rho_c=0.06$~fm$^{-3}$. Notice the very good agreement with the 
experimental $P_c$ and $\rho_c$ values from Ref.~\cite{elliott}. As a side remark, we 
also provide for this parametrization the compressibility factor ($Z_c=\frac{P_c}{\rho_c 
T_c}$), namely, $Z_c=0.31$. For the experimental values of critical parameters 
of Ref.~\cite{elliott}, $Z_c=0.29$.

\section{Summary and Conclusions} 
\label{summ-concl}

In the present work, we studied correlations between bulk quantities of symmetric nuclear matter at 
zero temperature and their critical parameters (CP), namely, $T_c$, $P_c$ and $\rho_c$ at finite 
temperature regime. We performed this analysis in the RMF model presenting nonlinear 
couplings in the scalar field $\sigma$ up to fourth order, here named the BB 
(Boguta-Bodmer~\cite{boguta}) model. The motivation for such an investigation comes from the 
results (correlations) presented by the nonrelativistic version of the BB model. As in previous 
works~\cite{bianca}, the correlations presented in the NRL model are reproduced also in the BB one, 
by imposing the same physical conditions needed to make the relationships arise. 

In order to explore in a fully analytical way the NRL model, we proceeded to include the temperature 
effects in the pressure equation of state by simply adding in Eq.~(\ref{nrl-press}) an ideal gas 
contribution. Such an approximation neglects quantum effects, but still reproduces qualitatively the 
van der Waals behavior of warm nuclear matter, as displayed in Fig.~\ref{nlpc-iso}. By imposing the 
critical conditions of Eq.~(\ref{conditions}) in the NRL model, we found that $\rho_c$, $T_c$ and 
$P_c$ are directly correlated with $\rho_o$, $B_o$, $m^*$ and $K_o$, as shown by 
Eqs.~(\ref{nrlrhoc})-(\ref{nrl-pc}). For fixed values of $\rho_o$ and $B_o$, the results pointed 
out to an increasing $K_o$ dependence of the CP. For $\rho_c$, this dependence is 
verified independently of the effective mass value, see Fig.~\ref{nrl-rhoc}. For $T_c$ and $P_c$, 
on the other hand, we verified a positive correlation with $K_o$ only for fixed values of 
$m^*$, see Fig.~\ref{nrl-tcpc}. Inspired by these results, we have calculated the CP of the $128$ BB 
parametrizations of Ref.~\cite{rmf}, looking for possible correlations with bulk quantities at 
$T=0$. We found a general trend of increasing values of the CP as $K_o$ increases, see 
Fig.~\ref{rmf-critical}. Such a trend is confirmed as clear correlations if we choose 
parametrizations in which the effective mass value is kept fixed, exactly as we have concluded 
in the NRL model case for $T_c$ and $P_c$. In Fig.~\ref{rmf-corr-Ko} we showed this analysis for BB 
parametrizations constructed with the ranges of $0.58\leqslant m^*\leqslant 0.64$ and $250\leqslant 
K_o \leqslant 315$~MeV. The former range~\cite{ls-splitting} ensures BB models presenting 
spin-orbit splittings within accepted experimental values, and the latter~\cite{stone} was recently 
proposed from a reanalysis of up-to-date data on isoscalar giant monopole resonance energies.

The comparison of our findings with previous correlations results of 
Refs.~\cite{kapusta,lattimer,natowitz,rios}, obtained from other hadronic models, pointed out to a 
qualitative agreement concerning the $T_c\times K_o$ correlation ($T_c$ is an increasing 
function 
of $K_o$). We also found a clear correlation between the CP and the effective mass if $K_o$ of each 
BB parametrization is kept fixed, see Fig.~\ref{rmf-corr-m}. For the sake of completeness, we also 
investigated the relationship of the CP with the saturation density and binding energy. Our results 
showed an increasing behavior of the CP with $\rho_o$, in qualitative agreement with the models of 
Refs.~\cite{kapusta,rios}. For the 
case of $B_o$, the BB model present $T_c\times B_o$ as an increasing function, while $P_c$ and 
$\rho_c$ exhibit a decreasing dependence, see Figs.~\ref{rmf-corr-rho} and~\ref{rmf-corr-b}, 
respectively.

A direct comparison of our findings for $T_c$ with experimental data collected from 
Refs.~\cite{natowitz,karn1,karn2,karn3,karn4,karn5,elliott} was performed in Fig.~\ref{tc-exp}. By 
constraining the BB parametrizations to present values of $0.58\leqslant m^*\leqslant 
0.64$~\cite{ls-splitting} and $250\leqslant K_o \leqslant 315$~MeV~\cite{stone}, we predicted 
critical temperatures of $14.2\leqslant T_c \leqslant 16.1$~MeV, lower than most of the experimental 
points, but compatible with four (of seven) of them within the error bars. By neglecting the 
restriction of effective mass, we could construct a BB parametrization presenting $T_c=18.3$~MeV, a 
value closer to the experimental data, including the more recent one of $T_c=17.9\pm 
0.4$~MeV from Ref.~\cite{elliott}. For such a parametrization, the effective mass is given by 
$m^*\sim 0.76$, a higher value than those from the range $0.58\leqslant m^*\leqslant 0.64$, 
obtained through an analysis of finite nuclei spin-orbit splittings. If we discard 
this constraint, we can use the correlation between $T_c$ and $m^*$ to predict new ranges of 
effective mass. For example, from Fig.~\ref{tc-exp}, we see that the range of $0.64\leqslant 
m^*\lesssim 0.76$ produces critical temperatures compatible with all experimental data. We remind 
the reader that this procedure indicates higher values for $m^*$, apparently not compatible with 
finite nuclei calculations of Ref.~\cite{ls-splitting}, but within an analysis of the BB model, i. 
e., a model with only mesonic self-interactions in the attractive scalar field $\sigma$. A more 
complete study, taking into account more sophisticated RMF models, such as that named ``type 4'' 
model in Ref.~\cite{rmf}, is needed in order to verify if the ranges of $m^*$ are kept and even 
to investigate the role played by the effective mass, and other bulk quantities, in possible 
correlations with the CP. We will address such study in a future work.

As a last remark, we verified in our study, the first one relating CP and bulk quantities at zero 
temperature of the RMF BB model, that BB parametrizations constrained to $0.58\leqslant m^*\leqslant 
0.64$ and $250\leqslant K_o \leqslant 315$~MeV present values of $0.19\leqslant P_c \leqslant 
0.27$~MeV/fm$^3$ and $\rho_c=0.05$~fm$^{-3}$, compatible with experimental values of 
$P_c=0.31\pm0.07$~MeV/fm$^3$ and $\rho_c=0.06\pm0.01$~fm$^{-3}$ of Ref.~\cite{elliott}. The 
theoretical values can be further improved if we relax the effective mass condition and choose, for 
instance, $m^*\sim 0.76$. In this case, we predicted $P_c=0.34$~MeV/fm$^3$ and 
$\rho_c=0.06$~fm$^{-3}$ for this particular BB parametrization ($m^*=0.7626$ and $K_o=315$~MeV).

\section*{Acknowledgements}

We thank the support from Conselho Nacional de Desenvolvimento Cient\'ifico e 
Tecnol\'ogico (CNPq) of Brazil, Funda\c{c}\~ao de Amparo \`a Pesquisa do Estado do Rio 
de Janeiro (FAPERJ) and Coordena\c c\~ao de Aperfei\c coamento de Pessoal de 
N\'ivel Superior (CAPES). M.~D. acknowledges support from FAPERJ, grant $\#$111.659/2014.

\appendix*

\section{Coefficients of Eqs.~(\ref{nrl-tc}) and (\ref{nrl-pc})}

The coefficients presented in the expression of $T_c$, Eq.~(\ref{nrl-tc}), of the NRL 
limit model, are given as follows,
\noindent
$a_n:$
\begin{align}
a_0 &= -\frac{\lambda}{3M},\quad a_1 = a_2 = a_3 = 2,\quad a_4 = 2\lambda
\end{align}
\begin{align}
a_5 = \frac{88\lambda}{5},\quad a_6 &= \frac{77\lambda}{5}.
\end{align}
$b_n:$
\begin{align}
b_0 &= 2E_{\mbox{\tiny F}}^{o},\quad
b_1 = -\frac{\left( 2M^{2} - 19E_{\mbox{\tiny F}}^{o}M +54E_{\mbox{\tiny F}}^{o2} 
\right)}{3\rho_{o}},
\end{align}
\begin{align}
b_2 &= -\frac{8M\left(M-6E_{\mbox{\tiny F}}^{o} \right)}{\rho_{o}^{2}},\quad
b_3 = \frac{2M\left(M-10E_{\mbox{\tiny F}}^{o} \right)}{\rho_{o}^{3}},
\end{align}
\begin{align}
b_4 &= \frac{2 \lambda \left( 9M^{2} - 70E_{\mbox{\tiny F}}^{o}M +120E_{\mbox{\tiny 
F}}^{o2} \right)}{9M E_{\mbox{\tiny F}}^{o} \rho_{o}},
\end{align}
\begin{align}
b_5 &= \frac{352\lambda \left(M-6E_{\mbox{\tiny F}}^{o} \right) }{45M\rho_{o}^{2}},\quad
b_6 = - \frac{77\lambda \left(M-10E_{\mbox{\tiny F}}^{o} \right)  }{30M\rho_{o}^{3}}.
\end{align}

For the critical pressure, $P_c$, showed in Eq.~(\ref{nrl-pc}) we have,
\noindent
$c_n:$
\begin{align}
c_0 &= -\frac{2\lambda}{15M},\quad c_1  = 1,\quad c_2  = \frac{4}{3},\quad 
c_3  = \frac{3}{2},
\end{align}
\begin{align}
c_4  = \frac{5\lambda}{4},\quad
c_5 &= \frac{64\lambda}{5},\quad
c_6  = \frac{121\lambda}{10}.
\end{align}
$d_n:$
\begin{align}
d_0 &= E_{\mbox{\tiny F}}^{o},\quad d_1  = b_1,\quad d_2  = \frac{4}{3}b_2, \quad
d_3  = \frac{3}{2}b_3,
\end{align}
\begin{align}
d_4 &= \frac{5}{4}b_4,\quad d_5  = \frac{16}{11}b_5,\quad d_6  = \frac{11}{7}b_6.
\end{align}

The remaining coefficients presented in both expression are,
\noindent
$t_{n1},t_{n2}:$
\begin{align}
t_{11} &= -\frac{M\left(M -4E_{\mbox{\tiny F}}^{o}\right)}{6\rho_{o}},
\end{align}
\begin{align}
t_{12} &= \frac{1}{\rho_o}\left[B_{o}\left( 9M^{2} - 48E_{\mbox{\tiny F}}^{o}M 
+18E_{\mbox{\tiny F}}^{o2} \right)\right. \nonumber\\
&+ \left. E_{\mbox{\tiny F}}^{o}M\left(4M-21E_{\mbox{\tiny F}}^{o}\right)\right],
\end{align}
\begin{align}
t_{21} &= \frac{M\left(M -3E_{\mbox{\tiny F}}^{o}\right)}{\rho_{o}^{2}},
\end{align}
\begin{align}
t_{22} &= -\frac{9B_{o}M\left(3M-13E_{\mbox{\tiny F}}^{o}\right) + 3E_{\mbox{\tiny 
F}}^{o}M\left( 5M-27E_{\mbox{\tiny F}}^{o} \right)}{\rho_{o}^{2}},
\end{align}
\begin{align}
t_{31} &= -\frac{M\left(M-2E_{\mbox{\tiny F}}^{o} \right)}{\rho_{o}^{3}},
\end{align}
\begin{align}
t_{32} &= \frac{6B_{o}M\left(3M-10E_{\mbox{\tiny F}}^{o}\right)
+6E_{\mbox{\tiny F}}^{o}M\left(M-6E_{\mbox{\tiny F}}^{o}\right)}{\rho_{o}^{3}},
\end{align}
\begin{align}
t_{41} &= \frac{M -6E_{\mbox{\tiny F}}^{o}}{9M\rho_{o}},
\end{align}
\begin{align}
t_{42} &= -\frac{B_{o}\left(6M-32E_{\mbox{\tiny F}}^{o}\right) 
-2M^{2} + 52E_{\mbox{\tiny F}}^{o}M/3 - 36E_{\mbox{\tiny F}}^{o2}}{M\rho_{o}},
\end{align}
\begin{align}
t_{51} &= -\frac{\left(M -3E_{\mbox{\tiny F}}^{o}\right)}{9M\rho_{o}^{2}},
\end{align}
\begin{align}
t_{52} &= \frac{B_{o}\left(3M-13E_{\mbox{\tiny F}}^{o}\right)
+E_{\mbox{\tiny F}}^{o}\left(5M/3-9E_{\mbox{\tiny F}}^{o} \right)}{M\rho_{o}^{2}},
\end{align}
\begin{align}
t_{61} &= \frac{M-2E_{\mbox{\tiny F}}^{o}}{6M\rho_{o}^{3}},
\end{align}
\begin{align}
t_{62} &= -\frac{ B_{o}\left(3M-10E_{\mbox{\tiny F}}^{o}\right)+ E_{\mbox{\tiny 
F}}^{o}\left(M-6E_{\mbox{\tiny F}}^{o} \right)}{M\rho_{o}^{3}}.
\end{align}

\end{document}